\documentclass[12pt,letterpaper]{article}
\usepackage{amsmath,amssymb,pgf,pgfarrows,pgfnodes,float,appendix, hyperref}
\usepackage{graphicx}
\usepackage{subfigure}
\usepackage[margin=0.9in]{geometry}

\newcommand{\be}{\begin{equation}}
\newcommand{\ee}{\end{equation}}
\newcommand{\bea}{\begin{eqnarray}}
\newcommand{\eea}{\end{eqnarray}}

\title{{\rm\footnotesize \qquad \qquad \qquad \qquad \qquad \ \qquad \qquad \qquad \ \ \ \ \ \                  RUNHETC-2020-04}\vskip.5in   Holographic Space-time and Quantum Information  }
\author{Tom Banks\\
Department of Physics and NHETC\\
Rutgers University, Piscataway, NJ 08854\\
E-mail: \href{mailto:banks@physics.rutgers.edu}{banks@physics.rutgers.edu}
\\
\\
}
\date{}
\begin{document}
\maketitle

\begin{abstract}
The formalism of Holographic Space-time (HST) is a translation of the principles of Lorentzian geometry into the language of quantum information. Intervals along time-like trajectories, and their associated causal diamonds, completely characterize a Lorentzian geometry. The Bekenstein-Hawking-Gibbons-'t Hooft-Jacobson-Fischler-Susskind-Bousso Covariant Entropy Principle, equates the logarithm of the dimension of the Hilbert space associated with a diamond to one quarter of the area of the diamond's holographic screen, measured in Planck units.  The most convincing argument for this principle is Jacobson's derivation of Einstein's equations as the hydrodynamic expression of this entropy law.  In that context, the null energy condition (NEC) is seen to be the analog of the local law of entropy increase.  The quantum version of Einstein's relativity principle is a set of constraints on the mutual quantum information shared by causal diamonds along different time-like trajectories.  The implementation of this constraint for trajectories in relative motion is the greatest unsolved problem in HST.
The other key feature of HST is its claim that, for non-negative cosmological constant or causal diamonds much smaller than the asymptotic radius of curvature for negative c.c., the degrees of freedom localized in the bulk of a diamond are constrained states of variables defined on the holographic screen.  This principle gives a simple explanation of otherwise puzzling features of BH entropy formulae, and resolves the firewall problem for black holes in Minkowski space.  It motivates a covariant version of the CKN\cite{ckn} bound on the regime of validity of quantum field theory (QFT) and a detailed picture of the way in which QFT emerges as an approximation to the exact theory.
\end{abstract}

\section{Introduction}

Every known human or computer language has the notion of time hard wired into every sentence.  One of Einstein's great insights was that this notion is relative.  Every information gathering system has its own {\it proper} time, and part of every physical theory must be a prescription for understanding the relations between the proper times of different systems.  His second great insight, that "the speed of light is finite", can be thought of as the definition of what we mean by space and space-time.  The region of space accessible to a system grows at a finite rate as a function of the proper time interval. The region of space-time accessible in a given time interval is called a causal diamond.  One can view time evolution along a time-like trajectory/set of nested causal diamonds as a foliation of the space-time manifold into space-like leaves. The variety of such trajectories means that this can be done in many different ways.  This led Einstein to formulate his theory of gravitation as a theory of the Lorentzian geometry of the space-time manifold.  

Lorentzian geometry can be recast as a theory of timelike trajectories and causal diamonds in a differentiable manifold.  A timelike trajectory is a one parameter choice of negative norm tangent vectors, and defines a positive number, the proper time interval, between any two points along the trajectory.   The causal diamond of a proper time interval is the set of all points that can be connected to both the past and future tips of the interval by timelike trajectories.  The boundary of any finite area\footnote{Here we're anticipating the definition of area that we are about to give.  In the limit of infinite area, the past and future halves of the diamond's boundary do not have to be joined differentiably. We also, by abuse of language, use the term {\it area} for  the $d - 2$ volume of a space-like slice of boundary of a causal diamond in $d$ dimensions. } diamond can be parametrized by two patches, with metrics
\begin{equation} ds_{\pm}^2 = du^{\pm} A_i^{\pm} (x,u^{\pm}) + g_{ij}^{\pm}  (x,u^{\pm}) . \end{equation}
The absolute maximum volume of the two Euclidean metrics $g_{ij}^{\pm} $, as a function of the null coordinates $u^{\pm}$ is called "the volume of the holographic screen of the diamond" , which we will distort to "the area of the diamond" as a shorthand. 

Although Jacobson did not use the language of causal diamonds, his seminal paper\cite{ted} showed that Einstein's gravitational equations follow as the hydrodynamics of a law that equates the "entropy" of a diamond to a linear function of its area.   The null energy condition (NEC) then follows from increase of entropy and is seen to be a thermodynamic statement, which will have fluctuation corrections.  Jacobson's derivation of Einstein's equations uses the frame of reference of a maximally accelerated trajectory to define energy.  Such a system has infinite temperature, which is the strongest argument that the entropy in the covariant entropy bound\cite{fsb} refers to the log of the dimension of the Hilbert space of the diamond\cite{boussobf}.  

The essence of Jacobson's argument, in the language of causal diamonds, is that the holographic screen\cite{bousso} of the diamond is, by its definition, a maximum of the area on the boundary.  Therefore, if we consider a pencil of null geodesics on the boundary of the diamond, approaching the holographic screen, then the Raychauduri equation can be linearized in the vicinity of the screen, and the increase of area can be written as 
\begin{equation} d A = R_{\mu\nu} k^{\mu} k^{\nu} d\lambda, \end{equation} where $\lambda$ is the affine parameter along the center of the pencil and $k^{\mu}$ is the null tangent vector.  By appropriate choice of diamond, $k^{\mu}$ can be any null vector in space-time.  The pair of future directed null trajectories following the boundary of the diamond past the holographic screen is the limit of a uniformly accelerated Unruh trajectory, with infinite Unruh temperature.  Defining energy to be the limit of $k^{\mu} k^{\nu} T_{\mu\nu}$, where $T$ is a covariantly conserved stress tensor, the equation $dE = T dS$, with $S = G_N A/4$, gives us exactly (in $4$ dimensions) \begin{equation} k^{\mu} k^{\nu} (R_{\mu\nu} - \frac{1}{2} g_{\mu\nu} R - 8\pi G_N T_{\mu\nu}) = 0 . \end{equation}  This is the content of Einstein's equation without the cosmological constant (c.c.).  We've thus derived the gravitational field equations as the hydrodynamic equations of the area law, and simultaneously shown that the c.c. is not a hydrodynamic energy density.

The area law for entropy is the clue for understanding locality/causality in a quantum theory of space-time.  Given a time-like trajectory, the causal diamonds of a nested series of proper time intervals partition the interior of the largest diamond into a sequence of quantum subsystems whose maximal entropy is non-decreasing as a function of the length of proper time.  Causality is the statement that the smaller subsystems remain unentangled with the rest of the degrees of freedom during the relevant proper time intervals.  This implies that time evolution is naturally viewed trajectory by trajectory\footnote{Jacobson's derivation of Einstein's equations from the first law of (local) thermodynamics uses the energy along a particular maximally accelerated trajectory and thus also points to a trajectory by trajectory view of time evolution.} and that the natural time slices inside a diamond must remain inside the diamond.  The Hamiltonian is perforce time dependent.  This can be viewed either as the quantum requirement of gradual entanglement of subsytems or, macroscopically, as the geometric requirement that time slices remain within a diamond.  As Milne\cite{milne} first appreciated, this kind of time slicing induces a redshifting of the "energies" of distant objects\footnote{Milne was of course incorrect in assuming that the observed cosmological redshift could be attributed entirely to this kinematic effect.} .

From a more philosophical point of view, what a formalism based on these ideas is saying, is that time is fundamental, but relative (trajectory dependent, many fingered), while space is an emergent concept describing a measure of the amount of quantum information required to describe a certain time interval.  The quantum analog of Einstein's principle of relativity then becomes apparent.  Consider a pair of causal diamonds along two different trajectories.   There is a maximal causal diamond in their intersection.  The Covariant Entropy Principle (CEP) assigns this diamond a Hilbert space of fixed dimension, which will always be smaller (geometry) than the dimensions of either intersecting diamond.  Each parent diamond is a quantum system with time dependent Hamiltonian and, given a choice of initial pure state, will assign a sequence of density matrices to the subsystem describing the intersection.  The Quantum Principle of Relativity (QPR) asserts that the two density matrices assigned by the parent diamonds have the same entanglement spectra. This constrains the choice of both the Hamiltonian and the initial state in each diamond.  We'll outline below the utility of this principle for trajectories at relative rest.  We have not yet found a model that implements the QPR for pairs of trajectories in relative motion.

While the CEP allows us to localize quantum information on the holographic screens of nested or intersecting diamonds, it does not give us a clear definition of a traditional localized excitation in the bulk of a given diamond.  The clue to bulk localization comes from two formulae in black hole physics.  The first is the entropy formula for Schwarzschild-de Sitter black holes.  The metric is
\begin{equation} ds^2 = - f(r) dt^2 + dr^2 / f(r) + r^2 d\Omega^2 , \end{equation} where \begin{equation} r f(r) = - (r - R_+) (r - R_-) (r + R_+ + R_-) , \end{equation} \begin{equation} R^2 = R_+^2 + R_-^2 + R_+ R_- , \ \ \ \ 2M R^2 = R_+ R_- (R_+ + R_-) . \end{equation} $R$ is the dS radius, and $M$ is the parameter that becomes the black hole mass in the $R \rightarrow \infty$ limit.  Everything is written in Planck units.  This formula shows that the introduction of any object with a long range Schwarzschild field, at rest on a timelike geodesic of dS space, creates an entropy deficit.  The CEP identifies this, when $M$ is small compared to the maximal black hole mass in dS space, as the deficit expected in a thermal ensemble with temperature $ T = (2\pi R)^{-1}$, the Gibbons-Hawking temperature\cite{gh}.   In fact one can demonstrate a similar entropy deficit effect in Minkowski space\cite{verlinde}, which suggests that the Minkowski vacuum is an infinite entropy ensemble.

The second hint that localized objects are constrained states of holographic variables comes from the ordinary formula for the entropy of a Minkowski black hole of mass $M$ 
when an additional small mass $m$ is dropped into it.  Despite the fact that the small mass is a very low entropy object, the final equilibrium state is a state of much higher entropy \begin{equation} \Delta S = 2\pi R_S m . \end{equation}   This indicates that before equilibration, the combined system lived in a much larger Hilbert space than that of the original black hole, but that the initial state had $ 2\pi R_S m$ frozen degrees of freedom. Inverting this process (by unitarity), we have a derivation of the Hawking temperature for emission of the system of mass $m$.   

If the Hamiltonian that equilibrates the system has a natural time scale $R_S$ and is a "fast scrambler"\cite{hpss}, then the infalling subsystem will remain isolated for a time of order $R_S {\rm ln}\ R_S$, and this is the basis for the resolution of the "firewall paradox"\cite{fw}.  Again, the principle operating 
here is that a localized state in the causal diamond formed by the horizon of the black hole of mass $m + M$ is a constrained state of the Hilbert space of that black hole. 
Another important feature we learn from this discussion is that the constraints must have the property that they isolate the degrees of freedom of the small system, from that of the black hole.  This immediately suggests that the degrees of freedom should form matrices, with a single trace Hamiltonian, and the constraints implying that off diagonal matrices connecting the "m-block" to the "M-block", vanish. 

Constraints can "propagate through a nested sequence of causal diamonds", giving a holographic interpretation of particle trajectories.  More properly we'll see that these should be thought of as jets of particles, including many soft gravitons whose number changes with time.  Indeed, we'll see that "particle" is a perturbative concept and jets are the fundamental scattering states in models of quantum gravity in Minkowski space. The trajectory of a jet, a quantum system with many states for fixed momentum, is a much more robust semi-classical object than a particle trajectory in quantum field theory.   

We'll see that the CEP, the QRP, the identification of particle jets as constraints, and the fast scrambling properties of black hole horizons give us a number of vital clues to the nature of a general theory of quantum gravity.  For example, the QRP enables us to tie together jet interactions (some number of jets enter the past boundary of a diamond and a possibly different number exit its future boundary) in different causal diamonds, obtaining a manifestly local, Feynman diagram like, description of transition amplitudes. The same formalism can describe the production and decay of high entropy meta-stable excitations with all of the qualitative properties of black holes.

\section{The Holographic Variables of Quantum Gravity}

The CEP implies that a finite area diamond corresponds to a finite dimensional Hilbert space.  The fact that the $U(D)$ GellMann matrices, which are closed under both commutation and anti-commutation, form a basis for all complex matrices shows us that this space is the fundamental representation of the super-algebra $SU(P|Q)$ for any integers $P,Q$ such that $P + Q = D$.  That is to say, fermionic variables are inevitable in any finite dimensional quantum system.  This remark ignores the constraint of spatial locality.   A discrete, spatially local system can be defined on the tensor product of finite dimensional Hilbert spaces sitting at the points of some graph, whose links define what we mean by nearest neighbor, next to nearest neighbor, {\it etc.} couplings.  Fermionic operators on the full Hilbert space will be non-local functions of the bosonic site variables.  In some cases\cite{jorwigkapustintb} a local theory of mutually commuting site variables, with a $Z_2$ gauge invariance, can be rewritten as a local theory of fermions, but this is not always the case.  

The fast scrambling property of black holes\cite{hpss} implies that the correct quantum theory cannot be local on the holographic screen of a diamond\footnote{In AdS space, for black holes larger than the radius of curvature, scrambling {\it is} ballistic on length scales larger than the AdS radius.  This is a consequence of the AdS/CFT correspondence.}.  Instead we will suggest that the Hamiltonian should be invariant under a finite dimensional approximation to the group of area preserving maps on the sphere.  The theory of fuzzy approximations to Euclidean geometries has a long history.  Traditionally it is viewed as the replacement of the algebra of smooth, or continuous, functions on the manifold by a finite dimensional non-abelian matrix algebra.  This can be developed in a systematic way for any manifold with a Kahler or symplectic structure. In \cite{tbjk} we proposed a different approach, inspired by Connes' insight about the connection between the Dirac operator and Riemannian geometry.  The Dirac operator on any spin manifold is an unbounded operator with spectrum symmetric around $0$ and compact inverse on the space of spinor sections orthogonal to its discrete zero modes.  Its eigenvalues are invariant under any symmetries of the manifold, and its zero mode spectrum encodes some of the topological properties.  The space of spinor bilinears is the space of all differential forms on the manifold, so appropriate products of spinor bilinears are proportional to its volume form and a Hamiltonian given by the integral over such products is invariant under area preserving maps.

\begin{figure}[h!]
\begin{center}
 \includegraphics[width=12cm]{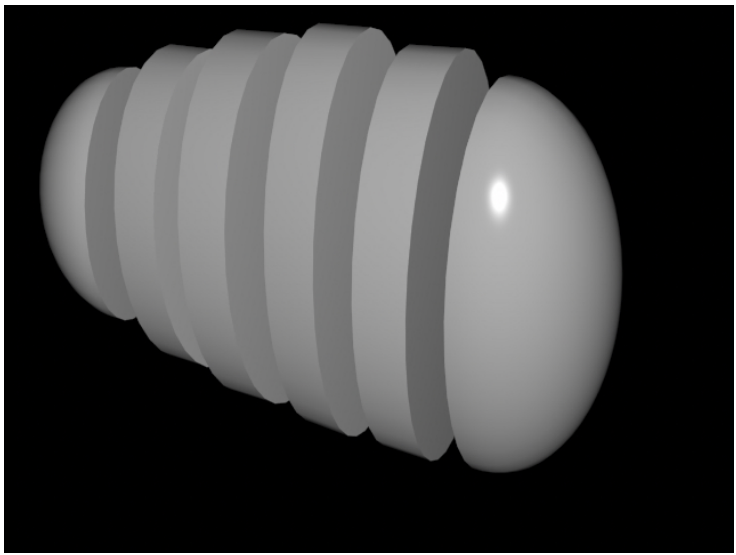}
\end{center}
\vspace{-1.4cm}
\caption{A monomial interaction. To simplify the picture, we did not attempt to illustrate the invariance under area preserving mappings, which could turn these regular slices into amoeba.\label{monomial}}
\end{figure}

We fuzzify the geometry by putting a symmetric eigenvalue cutoff $P > 0$ on the Dirac operator.  For large $P$, the eigenvalue degeneracy goes like $P^{d - 2}$ where $d$ is the space-time dimension, so if each eigensection is quantized in a finite dimensional Hilbert space of fixed dimension, then we get an area law for the maximal entropy.  On the $d - 2$ sphere, the counting of spinor spherical harmonics is exactly that of anti-symmetric $d - 2$ tensors with indices ranging from $1$ to $o(P)$\cite{tbjk}.  We can think of these as little area elements.  The matrix $M_i^j \equiv \psi_{i,a^{1} \ldots a^{d - 3}} \psi^{\dagger\ j,a_1 \ldots a_{d-3}} $ can be viewed as a $d - 3$ sphere "band" on the surface of a $d -  2$ sphere and the trace of a product of these matrices is a "line bundle" construction of the $d - 2$ sphere from a succession of such bands.  In plainer language, it's the picture of the $d - 2$ sphere as a succession of "thick" $d - 3$ spheres along a polar coordinate
Figure \ref{monomial}.

 \begin{equation} ds^2 = d\theta^2 + \sin^2 \theta d\Omega_{d - 3}^2 . \end{equation}  The bilinears in spinors are differential forms of varying degrees and the trace is the integral over $d - 2$ forms formed as wedge products of these elementary forms.  Any action constructed from sums of such single traces is invariant, in the formal continuous limit, under area preserving maps.  If we write 
 \begin{equation} H_{in} (t)  = \frac{1}{t} {\rm Tr}\ P(\frac{M_{t\times\ t}}{t^{d -3} }) , \end{equation} where $t$ is the proper time in a diamond, and $P$ is a polynomial of finite order whose coefficients are $t$ independent in the large $t$ limit, then the leading term in the energy scales like $P^{d - 3}$ as $t \rightarrow \infty$.  The gaps between an infinite number of low lying states and the ground state are $o(1/t)$ in this limit.  The CEP indicates that $P \sim t$ should be proportional to the radius of the sphere, in Planck units. The relation between the short wavelength cutoff on the sphere, and the proper time/area of the holographic screen, is a UV/IR correspondence, generalizing Maldacena's scale radius duality.
 
 The full Hamiltonian of HST is more complicated than this in constrained subspaces, which make the matrices block diagonal.   If we have a number of isolated blocks in a causal diamond of size $t$\footnote{At this point the reader should be prepared to understand "causal diamond" as "tensor factor of the Hilbert space which interacts only with itself over the time interval $[- t, t]$".}, then for each block of size $n_b$ we add 
 \begin{equation} H_{in}^{n_b} (t)  = k n_b^{d - 3} + \frac{1}{n_b} {\rm Tr}\ P(\frac{M_{n^b\times\ n^b}}{n_b^{d -3} }), \end{equation} where $k$ is a constant,which will be determined by the correct normalization of energies in the limits discussed below.  The first term is the "asymptotically conserved energy of the jet represented by the block", while the second term represents fragmentation of the jet into subject constituents.  The block COULD also represent an isolated black hole in the diamond and then the second term represents interactions on the black hole horizon.  Note that 
$H_{in}^{n_b} (t)$ is $t$ independent.  This is because it represents excitations localized near the trajectory, which have order $1$ energies in "Milne" coordinates.
There is also, of course, an $H_{out} (t)$ describing interactions of degrees of freedom outside the diamond.  We'll see below that this is determined by the QRP.

The commutation relations for these variables that are invariant under $SO(d - 1)$ are
\begin{equation} [\psi_{A} , \psi^{B} ]_+ = \delta_{A}^B  . \end{equation}  Here $A,B$ are $d - 2$ dimensional antisymmetrized multi-indices and the right hand side is the appropriately antisymmetrized Kronecker symbol.
These, and the Hamiltonian are invariant under the larger $U(t)$ group of unitaries, which can be interpreted as a fuzzy approximation to the group of area preserving maps.  It has many $SO(d - 1)$ subgroups under which the variables transform as a sum of spherical harmonics up to some maximal angular momentum (simply conjugate one $SO(d - 1)$ subgroup by a general element of $U(t)$).  

If the $\psi$ variables have another index $A$, apart from their $SO(d - 1)$ spinor label
we can try to view them as fuzzy spinors on a higher dimensional manifold, of the form ${\cal K} \otimes {\cal M}^{1, d - 1}$because the spinor bundle on a product manifold is a tensor product of spinors on the lower dimensional manifolds.  More research is needed to find restrictions on the commutation relations as a function of the $A$ label, which approach geometrically sensible rules, where the anti-commutator of two spinor 
generators involves forms integrated over closed cycles on a manifold, in the limit that the number of $A$ labels gets large.

Given the generators $\psi_{a^1 \ldots a^{d - 3}}$, we can define mutually commuting Pauli operators by multiplying each fermion by $( - 1)^{N - n}$ where $N$ is the total number operator and $n$ the number operator of that particular species. The bilinear
$$\psi_a^A \psi^{\dagger\ b}_A , $$ becomes $$\sigma_{-\ a}^A \sigma_{3\ a}^A \sigma_{3\ A}^b \sigma_{+\ A}^b . $$ Here $A$ is a $d - 3$ component index, representing an interface between two bands on the $d - 2$ sphere.  So we can "bosonize" these fermions without introducing any more non-locality than was present in the original Hamiltonian.  For models invariant under the fuzzy version of area preserving maps, fermionic variables are natural, invariant, and as local as a bosonic presentation of the same Hamiltonian.

An alternative view of the fermionic variables of HST comes from a proposal for generalized scattering theory for models of quantum gravity in Minkowski space\cite{cacb}.   Ordinary scattering theory for quantum field theories with a mass gap is based on the infinite set of asymptotically conserved LSZ currents
\begin{equation} j_{\mu}^f = i ( f_{\pm} \partial_{\mu}\phi - f_{\pm} \partial_{\mu} \phi ).  \end{equation}  Here $\phi$ is an interacting Hermitian Heisenberg field and $f_{\pm}$ normalizable positive or negative energy solutions of the Klein-Gordon equation with the physical particle mass.  Matrix elements of these currents in physical states are assumed to be conserved near the conformal boundary of Minkowski space, and this is true up to exponential corrections, to all orders in perturbation theory.  The physical Fock space is the representation space of the algebra of these currents and the Scattering operator interwines been the past and future bases.   This formalism breaks down for massless particles.

However, all particles that are massless for an entire range of couplings are associated with conserved currents.  For Goldstone bosons, where we can turn on a mass continuously, violating the current conservation law, the almost conserved current plays the role of the field $\partial_{\mu} \phi$ in massive scattering theory and it's plausible that the asymptotic Hilbert space is simply the representation space of the conserved current in the limit.  Similarly gauge and gravitational fields all have asymptotically conserved currents associated with them\cite{cacb}.  The stress tensor in gravitational models plays a special role because the joint spectrum of its asymptotically conserved currents is the momentum null cone, the Fourier dual of the conformal boundary.  All other conserved currents can be viewed as generalized functions on this cone.  That is, they are "quantized fields" on the momentum null cone.  We'll see that the reason for the scare quotes is that these operator valued generalized functions are not the conventional tempered distributions of axiomatic QFT.  The null cone is a singular manifold and conventional Wightman fields would not be well defined there.

More importantly, the behavior of black hole and cosmological quasi-normal modes indicates that the quantum systems living on horizons cannot have the approximate locality in angle that one would expect from even a lattice approximation to a conventional QFT, where the rigorous theorem of Lieb and Robinson proves that information transport over large distances is ballistic.  Instead one expects all of the degrees of freedom to be coupled together, without regard to metrical distance. Correspondingly the fields are not expected to satisfy differential equations.  The Hamiltonians we have written are not local, and are fast scramblers, because every fermionic variable is coupled to every other one by some term in the Hamiltonian.

The purpose of currents on the conformal boundary is to describe the flow of quantum numbers other than the momentum, at infinity.  Helicity or spin must be one of those quantum numbers, so we expect operators $H_i^{\pm} (P) $ carrying helicity out of/into the future/past null boundary and $\tilde{H}_i^{\pm}  (P)$ describing flows along the boundary. The two kinds of operators are related by space reflection, and only the tilde-free operators are needed to describe massless particles. 

When we retreat from the conformal boundary to a finite area causal diamond, $P$ must become a discrete label\footnote{We'll see later that it is an {\it emergent} label.} and the CEP implies that the Hilbert space on which the generators $H_i^{\pm} (P) $  and $\tilde{H}_i^{\pm}  (P)$ act must be finite dimensional, in which case we can always view the same space as generated by fermionic operators as above.  If we want the formalism to obey the spin-statistics theorem, then those fermions must carry half integer helicity and must, in the conformal boundary limit, take the form $Q_{\alpha}^I (P)$, $\tilde{Q}_{\alpha}^I (P)$, where $\gamma^a P_a Q^J(P) = \gamma^a \tilde{P}_a \tilde{Q} (P) = 0,$. Here $\tilde{P}$ is the space reflected null vector.   Note that these kinematic arguments do not imply that the model must be supersymmetric.  If all fermionic generators come in parity symmetric pairs, then the spin 3/2 particles that must accompany the graviton will be massive.  

The algebra of the left or right handed spinor generators is completely determined\cite{ags} by Lorentz invariance, cluster decomposition, and the absence of tensor charges in an interacting theory of particles.  It is
\begin{equation} [ Q^I_{\alpha} (P), Q^J_{\beta} (P^{\prime} )]_+ = \delta^{IJ} \delta (P\cdot P^{\prime}) \gamma^{\mu}_{\alpha\beta} M_{\mu} (P , P^{\prime}) . \end{equation}  $M_{\mu}$ is the smaller of the two parallel null vectors.  Note that the $P = 0$ generators anticommute with all the others.  There's a similar equation for the space-reflected generators.  The anti-commutation relations between the two sets of generators are not universal, and encode information about the masses of stable particles corresponding to branes wrapped around non-trivial cycles of a compact manifold.  As noted above, the detailed mathematics of the connection between finite dimensional super-algebras and the notion of smooth compact manifolds, has not yet been worked out.

\section{Time Dependent Hamiltonians and Error Correcting Codes}

The basic principles of HST imply that causality is implemented by gradually entangling new degrees of freedom in a larger causal diamond with the subset describing a smaller diamond contained in the original one.  We can ask where on the holographic screen of the larger diamond, the information about the smaller diamond is stored.  As long as the dynamics is invariant under (fuzzy) area preserving maps, this question has no meaning.   However, the constraints are a partial breaking of this symmetry.  The variables are labelled by spinor harmonic quantum numbers on the sphere, but there are an infinite number of ways of doing this, corresponding to the embeddings of $SO(d - 1)$ in the group of area preserving maps.  The constraints are interpreted in a way that mirrors a metric geometry on the sphere.  

For simplicity, let's work on $4$ dimensions.  In a large causal diamond with proper time $T$, we say that the physical state contains a localized jet on the past or future boundary if of order $ET$, with $E \ll T$,  of the variables $\psi_i^J$ vanish on that state.  Given the single trace nature of the interactions, this means that interactions between the variables $\psi_{[ij]}$ and the rest vanish on this state.  Here the small letters form an antisymmetric matrix with indices from $1$ to $E$, which can be organized into fuzzily localized spinor sections around some point $\Omega$.  We can think of the constraints as the vanishing of variables in an annulus surrounding a spherical cap, whose opening angle is determined by $E$.  More generally there will be multiple isolated subsets of degrees of degrees of freedom, which form $E_i \times E_i$ anti-symmetric matrices and are interpreted as belonging to spherical caps localized around different angles $\Omega_i$.  It can be shown\cite{bfsm} that $\sum E_i$ is an asymptotically conserved quantum number if the time dependent Hamiltonian has energy differences of order $1/T$.  Asymptotically, for large $T$, there will be a unique choice of rotation subgroup for which all of the jets form spherical caps localized around different points.  

To see this, note that we can always choose the $E_i \times (E_i - 1)$ components of a single jet's degrees of freedom to be localized functions around some particular point on the sphere, with localization radius $\sim 1/E_i$, by choosing a particular linear combination of spherical harmonics.  We can do the same for all the other jets, keeping them separated in angle, as long as $\sum E_i \ll T$, so that there are plenty of variables available to describe the empty angular regions with no jets.  In the limit $T \gg \sum E_i$ with $E_i \rightarrow \infty$ we can accomodate an arbitrary number of localized jets.  Note that the rotation subgroup we choose should also be used to organize the addition of degrees of freedom each time we increase the proper time by one Planck unit.  This corresponds to adding exactly one angular momentum multiplet of the chosen subgroup to the "in" Hilbert space for every Planck size tick of the clock.  It's also important to emphasize that the action of the rotations on the decoupled $(T - \sum E_i) \times (T - \sum E_i)$ block of the matrix is, in a sense, trivial since the interactions of these variables are invariant under are preserving maps that leave invariant the spherical caps at which the jets are located.  The ratio of the areas of those caps to the area of the sphere goes to zero in the limit.  Since the Hamiltonian of this large set of degrees of freedom goes to zero in the limit, they become topological degrees of freedom, sensitive only to the punctures on the sphere.  This is the HST description of the infinite dimensional space of arbitrarily soft massless particles that are present in any quantum theory of gravity in Minkowski space.

Thus, both rotation and time translation are asymptotic symmetries, as expected in a theory of gravity.  Note that the magnitude of the null momentum is an emergent quantity, proportional to $\sum E_i$.  To get a Lorentz invariant scattering operator we must take all $E_i$ to infinity at fixed ratio, keeping $\sum_i E_i  \ll T$.

As a consequence, the error correcting code\cite{ecc}\footnote{The connection between quantum error correction and bulk (AdS scale) locality was pointed out in the second paper of this reference, but was anticipated by the tensor network construction of Swingle. The general idea of error correction is to entangle the desired quantum information, with widely distributed degrees of freedom of a much larger system, so that erasing the part entangled with a few q-bits does not degrade the information.  The particular use of error correction in the AdS/CFT correspondence exploits/is limited by the locality of the boundary theory.  It is not appropriate for discussing horizons whose dynamics is invariant under area preserving maps.  HST claims to remedy this.} generated by the expansion of $H_{in} (t)$ to include more degrees of freedom, contains information that allows us to localize information about the constrained variables at angles.  Now however, consider the full evolution in the interval from $[- T, T]$.  The initial state satisfies constraints corresponding to incoming jets with energies $E_i$.  At some later negative time $ - t$ with $t < T$ there are two possibilities.  Either we inevitably reach a point where $\sum E_i \sim t$ or some of the constraints and decoupled degrees of freedom are not contained in the Hilbert space on which $H_{in} (t)$ acts.  In the former case the fast scrambling nature of the Hamiltonian implies that the constraints will be erased by the time one gets to the end of the time interval $[- t , t ]$.  The entire Hilbert space will be in equilibrium and we have a causal diamond with energy proportional to $\sum E_i$ filled with an isotropic system on its boundary, in equilibrium with entropy $(\sum E_i)^2$.  This system has all of the qualitative properties of a black hole.  Figure 2. shows cartoons of the two possibilities.
\begin{figure}[t!]
\begin{center}
\includegraphics[width=0.5\linewidth]{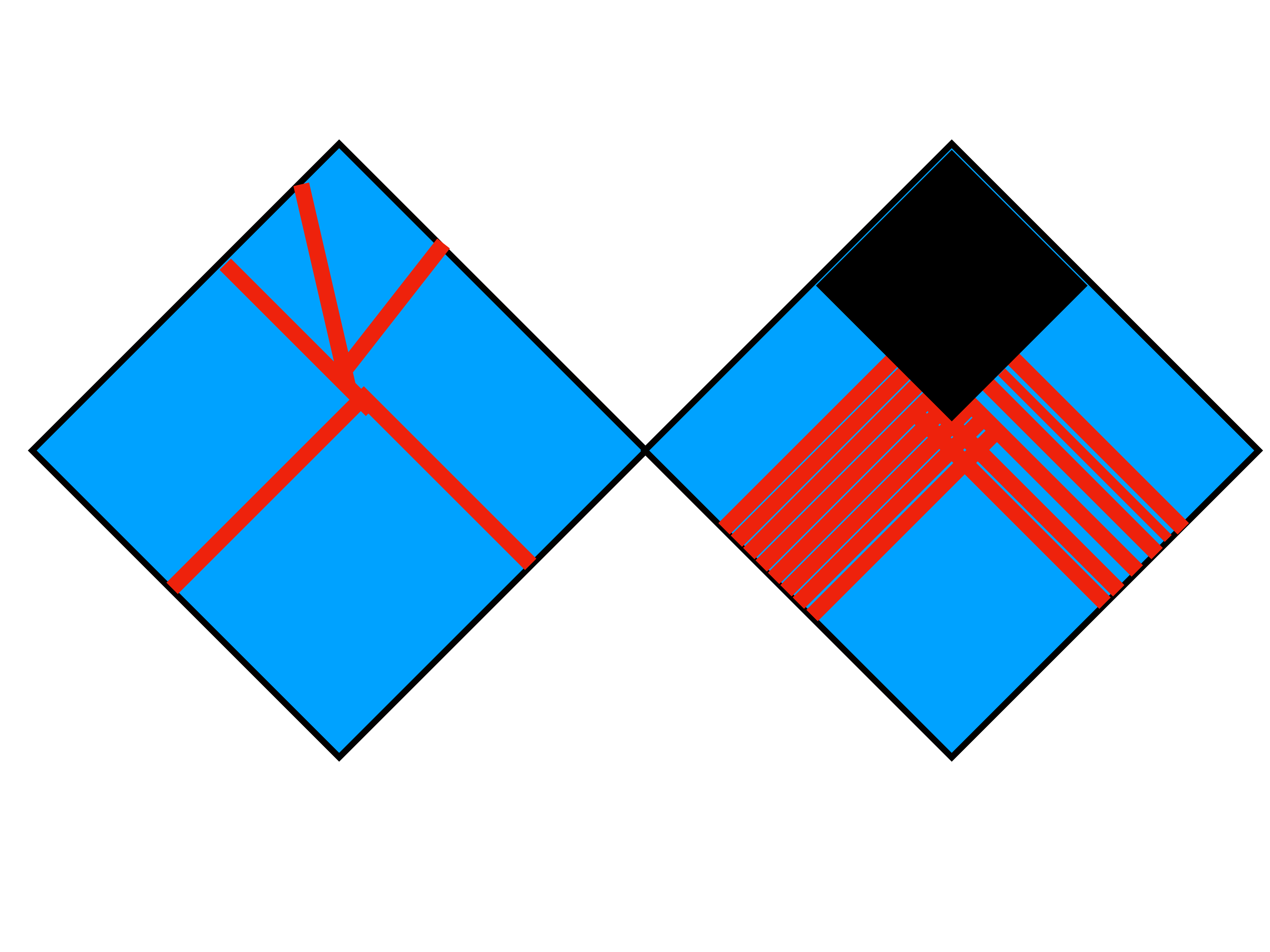}
\caption{The left figure shows a system with a number of constraints much smaller than the total number of degrees of freedom while the right one is what happens when the constrained subspace has entropy that is an order one fraction of the total. Red lines  denote jet degrees of freedom, each of which is surrounded on the past/future boundary of the diamond, by frozen degrees of freedom,indicated by erasure of the boundary.
The diamonds in these figures are finite, and the right hand picture does not include black hole evaporation.} 
\label{fig:twoposs}
\end{center}
\end{figure}

There are two different kinds of amplitudes where no black hole production occurs.  In the first, the total energy coming into the past boundary of the causal diamond $[ - T , T]$ is so small that $E^2$ is not a large entropy, {\it and} all of those constraints propagate into smaller diamonds along the same trajectory.  Then the future boundary of a diamond of proper time $t > E$ has a small number of constraints, which can be interpreted as jets of particles exiting that boundary.  On scales $ t = E $ the amplitude looks like a vertex in a Feynman diagram.  \vskip.2in

Another possibility is that the constraints proportional to the total incoming energy, which might be large, do not all propagate into small diamonds along the trajectory.  Here is where the overlap constraints of HST demonstrate the emergence of the concept of {\it space} in the HST formalism.  At time $t \ll T$, constraints that are not imposed on the Hilbert space of the $[ - t , t ]$ diamond, are imposed on its tensor complement in the $[ - T , T ]$ Hilbert space, which is acted upon by the Hamiltonian $H_{out} (t)$.   The structure of $H_{out} (t)$ is determined by the HST compatibility conditions, the QPR.  That is, given an assumed global structure of space-time, which is a dS space with $R \gg T\gg 1$ we can impose boundary conditions on causal diamonds with proper time $[ - T , T ]$ corresponding to $\sim E T$ constraints, with $E \ll T$, along time-like geodesics "at different spatial points in their common rest frame".  These are identical quantum systems, with the same sequence of time dependent Hamiltonians.  

Now consider, for a given initial state, the Hilbert spaces of these individual systems over time intervals $[- t + t_i , t_i + t]$ .   Let us first assume that the dynamics is such that in the large $T$ limit the proportionality constant $E$ in the number of constraints $ET + k$ is conserved. Call it the energy.  This is true for every individual trajectory.  When $E \sim t$ or greater, these cannot all be constraints on the "in" Hilbert spaces of the small causal diamonds.   Therefore, the energy must be divided between the in and out spaces.  The QPR implies that at any time, the "out" Hilbert space of a given diamond can be viewed as a tensor product of small Hilbert spaces corresponding to spatially separated diamonds.  If we take all of the $t_i$ equal, the translation of this statement into space-time language is that the energy $E$ is the sum of energies $E_i$ in each of the disjoint diamonds, with $E_i \ll t$ if we want to study a process in which no black holes are created.  The QPR implies that the "out" dynamics of any one diamond generates the same entanglement spectrum for the density matrices in each of the external diamonds, that is generated by the "in" dynamics of each of those diamonds.  Since we're studying geodesics in Minkowski space, the Hamiltonians are all assumed equal to each other, so the QPR is a constraint on initial states.\vskip.2in

\begin{figure}[h!]
\begin{center}
  \includegraphics[width=12cm]{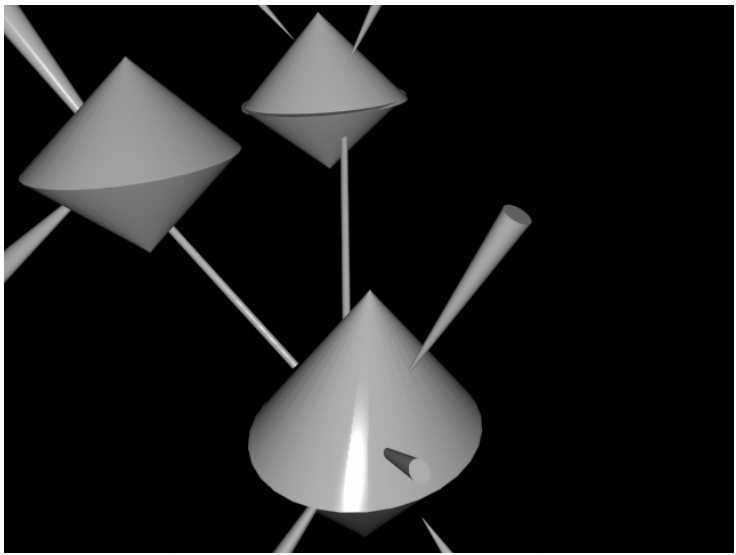}
\end{center}
\vspace{-1.4cm}
\caption{ Exchange diagram involving multiple jets.\label{exchange}}
\end{figure}

Now consider unequal $t_i$.   The final state constraints on the earliest diamond, can become part of the initial constraints on later diamonds, so we can draw a space-time picture of the amplitude that resembles a time ordered Feynman diagram Fig. \ref{exchange}
   Thus, these models reproduce the clustering structure of field theory amplitudes, which we usually derive from the postulates of QFT.  However, the HST formalism can also describe black hole formation and evaporation in a manner consistent with unitarity and causality.  So far, we have not found a model that gives Lorentz invariant scattering amplitudes. 

%\begin{figure}[t!]
%\begin{center}
%\includegraphics[width=0.5\linewidth]{qbitsandspacetime.pdf}
%\caption{The QPR requires us to tie together amplitudes in Feynman-like space-time diagrams.
%} 
%\label{fig:feynman}
%\end{center}
%\end{figure} 

We can close this circle by using QPR to finish the proof that the coefficient of $T$ in the number of asymptotic constraints {\it is} a conserved quantity.  If all of the asymptotic energy remains visible along a single time-like trajectory, conservation is a consequence of two facts.  The Hamiltonian in a causal diamond of proper time $t$ that is capable of removing constraints that prevent interaction between bulk and boundary DOF has eigenvalue differences of order $1/t$ and can only act to remove $o(1)$ constraints.  Furthermore, inside a causal diamond much of the "in" evolution over time $t$ acts on only a small part of the DOF.  For amplitudes in which the asymptotic energy divides into clusters localized in space-like separated diamonds, the QPR guarantees that the "out" Hamiltonian along a trajectory that currently sees only part of the energy, has the same effect as an "in" Hamiltonian acting on the individual energies $E_i$.

\subsection{Asymptotic Symmetries of HST}

We have just seen that time translation symmetry in HST arises as an asymptotic symmetry.  This is to be expected in a theory of gravity, but it's satisfying to see it arising from the quantum dynamics of an explicit model.  We've also shown that rotation symmetry arises asymptotically, acting only on the decoupled jet degrees of freedom.  It is asymptotic both because the organization of the DOF into spherical harmonics of a fixed rotation subgroup of the fuzzy volume preserving group depends on the asymptotic nesting of causal diamonds, and because rotations only act on the decoupled jets, which become truly independent of the rest of the system only in the limit of infinite proper time.

Spatial translation is more complicated.  Part of it is programmed into the construction of the model, by using the same sequence of time dependent Hamiltonians along each time-like geodesic of Minkowski space.  But the argument that scattering amplitudes are translation invariant comes from a combination of the QPR applied to asymptotically 
large diamonds along different geodesics, and the fact that jets decouple from "soft radiation" in the asymptotic limit.  Note that the QPR alone is insufficient because the overlap between two spacelike separated diamonds has parametrically smaller entropy than either of the diamonds in the infinite $T$ limit.   Thus, the QPR says only that the density matrix on the overlap is maximally uncertain, subject to the constraints.  The QPR also says that the angular location of the constraints seen along one trajectory, should look like a spatial translation of the angular locations as seen from the second.  Since the bulk of the variables decouple and freeze in the large $T$ limit, this suggests there is an identical Hilbert space, consisting of jets only, along the two relatively translated trajectories, and that the density matrices in that Hilbert space are related by a unitary transformation.  Thus, spatial translation is an asymptotic symmetry as well.

\section{Conclusions}

The basic principles of the HST formalism are the CEP and the implementation of causality by the unfolding entanglement of degrees of freedom in a nested set of causal diamonds.  The unitarity of the entangling map implies that this has the properties of an error correcting code.  The fact that the fundamental variables are the fermionic generators in the fundamental representation of $SU(K | L )$\footnote{Equivalently, they are canonical fermions with constraints.} follows from the finite dimension of the Hilbert space and the fact that dynamics is invariant under fuzzy area preserving maps of the holographic screen.  Area preserving invariance is valid for non-negative c.c. and for proper times sufficiently small compared to the AdS radius for negative c.c. , as seen from the behavior of black hole quasi-normal modes.  The fact that the fermionic operators must transform like asymptotic spinors, follows from the {\it presumed} Lorentz or AdS invariance of the boundary amplitudes, and the usual connection between spin and statistics.  

For negative c.c. and proper times of order the AdS radius and larger, quasi-normal mode analysis suggests and AdS/CFT dictates, that propagation is ballistic on the screen, on distance scales in the bulk larger than the AdS radius.  In order to make this compatible with the CEP for finite area causal diamonds with proper time approaching the AdS radius, we have to make a lattice field theory out of the fermionic variables and invoke the Lieb-Robinson bound.  The unfolding entanglement map that implements propagation in proper time is then the inverse of a Tensor Network Renormalization Group map in the sense of Evenbly and Vidal\cite{ev}.

Quantum information about small regions in the bulk is spread non-locally on the holographic screen, in a manner similar to that found in Error Correcting Codes.  Omission of the information in a small area region of a big screen does not destroy the data about the small causal diamond, because the entanglement of the small diamond's variables with those of the large screen is shared uniformly among the large screen variables.  The rate of spread over the area (which by the CEP is essentially the number of q-bits on the screen) undergoes a sort of phase transition in asymptotically AdS spaces, when the proper time in the diamond approaches the critical value (of order the AdS radius) at which the area of the diamond goes to infinity.  Prior to that regime, the scrambling is "fast" and the information is homogenized on the sphere in a time of order the radius of the sphere times the logarithm of the total number of q-bits.  As the proper time approaches the critical value, information scrambling is "fast" only over an area of order the $ d - 2$ power of the AdS radius.  If we imagine a collection of local probes, separated by distances on the holographic screen which are of order the AdS radius, communication between those probes is ballistic.  The system thus behaves like a lattice approximation to a quantum field theory.

In \cite{hstads} the authors conjectured that in the regime of the transition the HST Hamiltonian was the inverse of a Tensor Network Renormalization Group (TNRG) transformation\cite{ev}.  TNRG transformations disentangle the degrees of freedom of a fine grained lattice field theory at its critical point, producing a Hamiltonian on a more coarse grained lattice.  It's been shown by numerical analysis of simple one dimensional critical systems that the coarse grained Hamiltonians have a spectrum equal to that of the low lying levels of the radial quantization Hamiltonian of the CFT that describes the critical point.  Radial quantization always picks out a particular element of the conformal group as the Hamiltonian, and this is tied to a particular timelike geodesic, making an explicit connection with HST. This is a direct implementation of the scale/radius duality of Maldacena.  The TNRG can probably be improved via the technology of\cite{happy}, which constructs tensor networks invariant under discrete subgroups of the conformal group.

The HST formalism adds an extra element to the TNRG formulation of asymptotically AdS dynamics.  The TNRG Hamiltonian corresponding to some fixed proper time is conventionally defined to act as the unit operator on the tensor complement of the small Hilbert space corresponding to that causal diamond.  In HST language, it is $H_{in} (t) + 1$ for the last time slice in that diamond.  In HST, we have a Hamiltonian $H_{in} (t) + H_{out} (t)$ where $H_{out} (t)$ acts on the tensor complement.  In HST $H_{out} (t)$ is supposed to be determined by the consistency conditions with time evolution along other timelike trajectories.  In AdS space, all time-like geodesics are related by elements of the conformal group, so at least some of these consistency conditions are guaranteed asymptotically by the restoration of conformal symmetry implicit in any RG transformation at a fixed point.  It's possible that the conditions for finite diamonds and accelerated trajectories add further constraints.  

This implementation of HST has implications for the CFT description of diamonds of size much smaller than the AdS radius.  The work of Evenbly and Vidal shows that the finite dimensional Hamiltonians of the TNRG can be chosen to have the same spectrum as the low lying part of the exact CFT spectrum.  This is a very explicit implementation of Maldacena's scale-radius duality.  However, in the HST model, this correspondence breaks down as the proper time in the diamond is taken smaller than the AdS radius.  Instead of a lattice field theory we have a highly degenerate Hamiltonian with area preserving map invariance and fast scrambling.

One can argue that this disturbing disconnection must be a property of the AdS/CFT correspondence without any reference to HST.  Consider the causal diamond along a particular timelike geodesic in AdS space with proper time interval much smaller than the AdS radius.  Now consider the Witten diagrams for a correlation function of a finite number of operators on $R \times S^{d - 2}$.  The vertices of the diagrams are integrated over the entire AdS space, whose spatial volume on global time slices is all concentrated near the boundary.  Thus, the contribution to that correlation function from the causal diamond is very small, but non-zero.  The probability that the interactions take place within that diamond is small and is dominated by contributions from the boundary of the diamond.   As a consequence, if there is a notion of measurements localized in the diamond, they must register a state that is close to "empty Minkowski space" , with deviations concentrated near the boundary of the diamond.  This is consistent with the fact that the "energy" of such boundary states, in any coordinate system with spacelike slices localized in the diamond, will be very small.  This is an AdS/CFT argument that the "vacuum" of the approximately Minkowski region is a nearly degenerate ensemble rather than a single pure state.

The standard derivation\cite{psetal} of Minkowski amplitudes from CFT correlators illustrates the same principle.  All of the work on this subject has concentrated on showing that specially prepared 4-point functions converge to tree level 4 point scattering amplitudes\footnote{There are all sorts of caveats to this statement, particularly to its extension to higher point amplitudes which require us to study states localized in the large compact directions of $AdS_d \times {\cal K}$, but they've been discussed elsewhere.} .   But now consider a $4 + n$ point function with $n$ operators 
not constrained to focus on a particular "arena" causal diamond.  This gives a slightly different amplitude in Witten diagrams, but causes only small changes to the contribution from the bulk of the arena.  The obvious interpretation of these correlators is as a superposition of amplitudes for "n gravitons" and other soft massless particles to be injected into or emitted from the arena.   The full Minkowski scattering operator for "2 to 2" scattering includes all of these processes.

Note that in the discussion above there was no particular restriction to which $n$ operators were inserted.  We used only the fact that in connected diagrams involving all $4 + n$ particles, most of the vertices were integrated over all of AdS space.  Thus, the precise definition of the limiting Hilbert space on which the Minkowski scattering operator acts requires us to find a basis of states that can reproduce all of these amplitudes.   This is an unsolved problem in AdS/CFT.

If we accept the CEP for finite area diamonds in AdS space, Page's theorem\cite{page} implies that the empty Minkowski vacuum is a maximally uncertain density matrix.  In the large radius limit,  the ensemble consists of all states on the boundary of the arena that can be created by Witten diagrams with little weight in the arena causal diamond.  It is maximally uncertain because the number of possible Witten diagram states is much larger than the number of states that the CEP allows in the diamond.  States that correspond to scattering in the arena must then be constrained states of this ensemble.  That is, we've recovered the picture of localized excitations as constrained states of an ensemble of low energy excitations on the horizon.

In summary, HST treats time as fundamental, discrete and relative.  Space-time is an emergent phenomenon, measuring the amount of quantum information accessible to an information gathering system on a given timelike trajectory in fixed intervals of proper time.  This gives us a quantum definition of causal diamonds.  Causal propagation is an error correcting code by which quantum information about events in a small diamond is entangled with the states of a larger diamond containing it.  The information is spread rapidly over the holoscreen of the larger diamond, homogeneously for non-negative c.c., or for holoscreen sizes $\ll$ the AdS radius for negative c.c. .  On proper time scales of order the AdS radius, information is concentrated in the nodes of a tensor network and spreads ballistically over the network.  The spatial size of the nodes is of order AdS radius, as is the spacing between them.

The principle, valid for small enough diamonds with any c.c. and any diamond with non-negative c.c., that bulk localized states are constrained states of boundary DOF, with bulk energy proportional to the number of constrained q-bits, 
 explains most of the qualitative features of black hole and cosmological horizons and eliminates the firewall paradox.  The HST model gives a very explicit picture of the transition between an ordinary scattering event and black hole formation.  Black holes form when the energy, {\it the number of constraints}, entering into the past boundary of a causal diamond, creates a state so atypical that the fast scrambling Hamiltonian eliminates those constraints before the energy exits the diamond.  Thus, the boundary of validity of effective field theory ideas is an entropy bound.  The bulk localized entropy must be less that $S^{3/4}$ in four dimensions in order to avoid black hole formation.  This is a covariant version of a bound conjectured by the authors of\cite{ckn}.  Their bound was based on trying to understand the failure of field theory to compute the cosmological constant.  In HST, the c.c. is an input, but it is correct that the expectation value of the Hamiltonian in dS space does scale like the integral of the c.c. over the spatial volume of the static patch.  
 
 The resolution of the firewall paradox for non-negative c.c. is also completely entropic and can be understood without any of the details of the HST formalism.  The fact that black holes in these space-times have negative specific heat implies that the state just prior to the event we call "dropping a low entropy system onto a black hole", has a huge entropy deficit relative to the equilibrated black hole of slightly higher mass.  If there is any notion at all of a finite dimensional Hilbert space associated with the equilibrated system, then the pre-equilibrium state must be a low entropy constrained state in that Hilbert space.  Combining this with the fact that dynamics on the horizon has a natural time scale of order the black hole radius, and the natural conjecture that the frozen degrees of freedom mediate the interactions between the low entropy system and the original black hole horizon\footnote{In the HST model the constraints are precisely those that imply the off diagonal matrix elements of the matrix connecting the "black hole block" and the "infalling block", vanish, so that the two blocks don't interact.}, we find that there will be a time of order $R_S {\rm ln}\ R_S$
 during which the infalling system behaves as if the black hole were not there.  This is the "temporary" definition of "the smooth part of the black hole interior".  After the scrambling time the last phrase in scare quotes has no meaning, but a new infalling system will create its own interior.  Firewalls are a consequence of insisting on an invalid quantum field theory picture of quantum states near the horizon.
 \vskip.3in
\begin{center}
{\bf Acknowledgments }\\
I'd like to thank W. Fischler for years of joint work that went into formulating the HST formalism described here.  This work was supported in part by the U.S. Department of Energy under grant DE-SC0010008.
\end{center}

\end{document}